\documentclass[conference]{IEEEtran}
\pdfoutput=1

\usepackage{algorithm}
\usepackage{algpseudocode}

\algrenewcommand\alglinenumber[1]{{\sf\footnotesize#1}}
\algrenewcommand\algorithmicrequire{\textbf{Input:}}
\algrenewcommand\algorithmicensure{\textbf{Output:}}

\newcommand{\balsamjob}{\texttt{BalsamJob}}
\newcommand{\balsamjobs}{\texttt{BalsamJobs}}

\usepackage{enumerate}
\usepackage{cite}
\usepackage{times}
\usepackage{url}
\usepackage[hidelinks]{hyperref}
\usepackage{hyperref}
\usepackage{hypernat}
\usepackage{graphicx}
\usepackage{setspace}
\usepackage{subfigure}
\usepackage{mdwlist}
\usepackage{amsmath}
\usepackage{amssymb}
\usepackage{comment}
\usepackage{tabularx}
\usepackage{xspace}
\usepackage{color}
\usepackage[frozencache, cachedir=.]{minted}

\usepackage{hyperref}
\usepackage{cleveref}

\crefname{listing}{source code}{source codes}

\begin{document}

\title{Balsam: Automated Scheduling and Execution of Dynamic, Data-Intensive HPC Workflows}

\author{
\IEEEauthorblockN{Michael A. Salim, Thomas D. Uram, J. Taylor Childers, \\
Prasanna Balaprakash, Venkatram Vishwanath, and Michael E. Papka}
\IEEEauthorblockA{Leadership Computing Facility \\
Argonne National Laboratory\\
Lemont, IL 60439\\
Email: msalim\url{@anl.gov}}
}

%

\maketitle


\begin{abstract}
    We introduce the Balsam service to manage high-throughput task scheduling
    and execution on supercomputing systems. Balsam allows users to populate a
    task database with a variety of tasks ranging from simple independent tasks to dynamic multi-task
    workflows. With abstractions for the local resource scheduler and MPI
    environment, Balsam dynamically packages tasks into ensemble jobs and
    manages their scheduling lifecycle. The ensembles execute in a pilot
    ``launcher'' which (i) ensures concurrent, load-balanced execution of
    arbitrary serial and parallel programs with heterogeneous processor
    requirements, (ii) requires no modification of user applications, (iii) is
    tolerant of task-level faults and provides several options for error
    recovery, (iv) stores provenance data (e.g task history, error logs) in the
    database, (v) supports dynamic workflows, in which tasks are created or
    killed at runtime. Here, we present the design and Python implementation of
    the Balsam service and launcher. The efficacy of this system is illustrated
    using two case studies: hyperparameter optimization of deep neural networks,
    and high-throughput single-point quantum chemistry calculations. We find
    that the unique combination of flexible job-packing and automated
    scheduling with dynamic (pilot-managed) execution facilitates excellent
    resource utilization. The scripting overheads typically needed to manage
    resources and launch workflows on supercomputers are substantially reduced,
    accelerating workflow development and execution.
\end{abstract}

\IEEEpeerreviewmaketitle

\section{Introduction}
\label{sec:intro}

The rising value of data science methods across scientific domains is causing a
shift in the HPC landscape: there is a growing focus on \emph{data-intensive}
applications that produce or consume large datasets for analysis by modern
machine learning or broader statistical techniques.  Quite often, these
data-intensive applications join a variety of simulation programs and analysis
pipelines into large and dynamic workflows, comprising interdependent single-
and multi-node tasks.  Many instances of a given task may run concurrently in
large ensembles, spanning the requisite input space to generate the target
data.  With exascale computing on the horizon, data-intensive and
high-throughput computing workflows increasingly demand a strategy for
managing the sheer number of jobs in a campaign.

Several workflow management systems (WMS) directly address this problem
(Section \ref{sec:related}).  Although many solutions offer a powerful
framework for expressing complex workflows, \textbf{the tools for scheduling
and executing workflows on supercomputers are lacking in flexibility and
ease-of-use}. Interfacing with the local resource scheduler is an important
facet of supercomputing, yet even in sophisticated management solutions, the
burden often lies on the user to assign workflow resources and schedule batch
jobs. Automated queue submission is either rigid with respect to job queueing
parameters, or it requires configuring resource management carefully on a
\emph{per-workflow} (or even \emph{per-task}) basis.  Depending on the choice
of WMS, the user is also faced with one or more limitations in execution of
multi-program, multi-data (MPMD) jobs. Large ensembles of parallel tasks
coordinated by MPI have limited fault tolerance and can require significant
source code modifications.  Alternatives are usually lacking in dynamicity:
tasks of varying sizes and durations cannot be launched one-after-another to
optimally fill the allocated compute nodes and time. 

Motivated by these difficulties, we developed the Balsam HPC service to support
automated task scheduling and execution of dynamic workflows. Like other WMS
with notions of provenance and a centralized database, Balsam provides command
line and Python interfaces to control a task database. Users need only specify
the resources needed at a task level; the scheduling process then
\emph{packages} tasks and queues them for execution in dynamically-sized
ensemble jobs. These ensembles are sized to leverage a configurable scheduling policy.
This is essential to high-throughput workflows in \emph{capability}
supercomputing environments, because local scheduler policies typically favor large
(several hundreds of nodes) jobs and may even preclude scheduling jobs that are
otherwise routine in smaller HPC clusters.

A guiding principle in the Balsam design was ease-of-use: with minimal
configuration, the service provides a system-wide, multi-workflow solution for
scheduling and execution of jobs. Applications require no modification to run
inside Balsam. Simple command line interfaces inspired by the \texttt{conda}
package manager facilitate context switching, creating, and manipulating
workflows. The Python API (Section \ref{sec:dagdefining}) enables efficient
set-up of large job campaigns. For more complex
scenarios, the Django \cite{django} ORM underlying Balsam is accessible to users
for highly expressive queries and dynamic workflow modifications.

Execution is handled by the Balsam \emph{launcher}, a dynamic pilot
application that is automatically dispatched to run in batch job queues. Balsam
workflows are naturally fault-tolerant and support massive, multi-user job
campaigns and concurrency at leadership supercomputing scales.  In addition to
entirely pre-scheduled workloads, the launcher carries out dynamic workflows,
allowing tasks to spawn additonal tasks, or for tasks to be killed at runtime
and replaced on-the-fly with newly available work in the Balsam database.

In the following section, we compare Balsam with a few other WMS tools in the
context of the aforementioned challenges.  We then provide a high-level
overview of Balsam functionality and some details on the implementation, rooted
in Python and the Django \cite{django} framework. Two case studies are
presented: the first describes the DeepHyper \cite{deephyper} package for
hyperparameter optimization of deep neural networks, illustrating the coupling
with Balsam's Python API and support for dynamic task creation/killing.  The
second application showcases the utility of Balsam for executing a large batch
of parallel quantum chemistry single-point molecular calculations.

\section{Related Work\label{sec:related}}

The current work is an extension of the Balsam component of an HPC edge service
\cite{hpc-edge17} facilitating remote job submission and workflow management
for High Energy Physics experimental collaborations.  Balsam was designed to
provide a uniform interface to HPC resource schedulers, enabling remote job
submission from a centralized management service (\emph{Argo}).  Previously,
Balsam jobs were received on a message queue and individually throttled into the
local scheduler. Each queue submission corresponded to an independent task, and
there was no support for packing tasks into ensemble jobs or dynamic workflows
in Balsam. The Argo service was responsible for coordinating workflows and
mediating data transfer to and from Balsam sites. 

This work represents a significant overhaul of the Balsam architecture to
support ensemble job packing and dynamic workflows, while de-emphasizing the
facilities for remote job submission and multi-site execution. Here, we
highlight a few relevant WMS tools and discuss their strengths and limitations
in the context of the current work.

\subsection{FireWorks}

At a high level, the Balsam database and launcher are functionally analogous to
the FireWorks (FWS) \cite{fireworks15} infrastructure, consisting of a
centralized MongoDB \cite{mongodb} datastore (\emph{LaunchPad}) and clients
(\emph{FireWorkers}) that pull tasks for execution on the backend. 
Both FWS and Balsam leverage the database for job provenance, fault
tolerance / restartability, and dynamic workflows (wherein tasks alter the
database at runtime). 
In FWS, users can configure a \emph{QueueLauncher} service to automate
submission of a templated batch script to job queues over time.  Since queue
parameters are fixed in advance, this approach is suboptimal if the number or
size of runnable tasks varies significantly over time. Opportunities can be
missed to pack more tasks into larger ensembles as a workload grows, and these
jobs are typically prioritized in capability supercomputing systems. Alternatively,
submitting jobs from a workflow on a per-task basis precludes efficient job packing
and hampers throughput on systems with strict limitations on the total number of queued
jobs.

The Balsam service overcomes these limitations by sizing and queueing ensemble
jobs in an \emph{elastic} fashion, matching the net demands of a user's
workload with appropriately sized queue submissions.  The Balsam launcher
naturally handles scheduling departures (e.g. due to faults) by
dynamically assigning tasks to idle resources to maintain load balance. Another
key advantage of the launcher is that arbitrary MPMD workloads, containing MPI or
single-node tasks of varying sizes, are automatically run across the available nodes
without user intervention or any modification of applications.

\subsection{Swift/T and Parsl}
A number of related WMS may be classified by the ability to express a workflow
and its associated task graph directly \emph{as code}.  Swift \cite{swift} is a
compiled dataflow language, and Swift/T programs run statements concurrently
under a monolithic MPI application using the Turbine \cite{turbine} and ADLB
\cite{adlb} runtime libaries.  Parsl \cite{parsl} is a Python library that uses
Swift/T as an execution backend for supercomputing environments, providing
Python decorators to \emph{annotate} existing codes for asynchronous,
data-parallel execution. These WMS do not have a notion of a centralized task
database and are less appropriate than Balsam or FWS for long-running campaigns
requiring provenance data and cross-workflow schedulability.

Some difficulties arise when Swift/T workflows include parallel MPI tasks.
First, porting MPI programs to run in the Turbine environment requires
nontrivial development effort: the parallel program must be restructured as a
library with a callable entry function. Moreover, the tight coupling of all
tasks under one MPI communicator hampers fault-tolerance. When a single task
experiences a fault, all other processes in the workflow are forced to abort.
There are efforts underway to improve the ease-of-use and stability of MPI
programs launched within a larger MPI workflow \cite{comm-launch}. For the time
being, Balsam handles ensembles of parallel tasks by issuing a separate
\emph{mpirun} (or equivalent) command run for each task. While this approach is
less scalable due to the process demands placed on the head node,
fault-tolerance is significantly improved and the development overheads for
MPMD workflows with Balsam are reduced to none. We emphasize that Balsam
handles ensembles of thousands of single-node tasks under a single
\emph{mpirun}, and parallel tasks requiring just tens of nodes are already
large enough for capability-scale ensemble jobs that don't overload the head
node.  In scenarios where more than a few hundred small MPI jobs must run
concurrently, it is possible to use Balsam in conjunction with methods like
Cram \cite{cram} (discussed below) for finer-grained job packing.

\subsection{Cram}
Cram \cite{cram} is a tool that enables running many MPI jobs concurently as
one large application.  It is not intended as a complete WMS; instead, the Cram
library is simply linked against user codes to enable running an application in
ensemble mode without external services. This approach is a powerful workaround
for operating system limitations on the number of concurrent MPI processes, and
is interoperable with WMS like Balsam. Cram alone does not provide a complete
solution for high-throughput task execution, because it is a single program
parallelization scheme (SPMD), and it requires upfront definition of the tasks
in a Cram job file. This static behavior also precludes running successive jobs
in time; all tasks run concurrently and the ensemble takes as long as the
slowest instance. Moreover, since all Cram tasks run under the same
\texttt{MPI\_COMM\_WORLD}, if any task fails, the entire ensemble job will abort.
It is conceivable to use Balsam for dynamic and fault-tolerant execution of
many small Cram jobs. This approach would lessen the burden of MPI application startup
on the head node through fine-grained packing at the Cram level.

\section{Balsam Design and Features}

Here, we interleave a discussion of Balsam internals and interfaces with a
tutorial walkthrough of Balsam usage.  This section describes how to
define independent tasks or tasks with dependencies that form a directed
acyclic graph (DAG) workflow.  We will walk through available modes of job execution
and/or automated scheduling, and discuss how dynamic workflows can be
constructed to add, modify, or kill tasks at runtime according to user-programmed logic.
Along the way, various methods to examine Balsam for provenance and performance
data are illustrated.

\subsection{Overview of Balsam Components}
Balsam has three core components: the task database, launcher, and service.
The \textbf{task database} is the central Balsam datastore, with a \balsamjob{}
table holding one row for each task added to Balsam.  Each task corresponds to
a single run of a particular application and provides the necessary arguments,
environment variables, and so on.\footnote{We use \emph{task} to refer to a
single application instance and \balsamjob{} to refer to the corresponding
Python object model.}

The Balsam \textbf{launcher} runs
inside a compute resource allocation, \emph{pulls} tasks from the database, and
manages their concurrent execution.  The launcher automatically determines
available compute resources from the job environment and supports heterogeneous
workloads of arbitrary serial and parallel programs. Each task runs in an isolated
working directory; as execution proceeds, the
database is updated with the task state and other provenance data.
After setting up a workflow, users may invoke the \texttt{balsam launcher}
command directly in an interactive shell or job script. By submitting launcher
jobs to a queue, one has direct control over resource allocation and
scheduling. 

\begin{figure}[!t]
    \centering
    \includegraphics[width=3.0in]{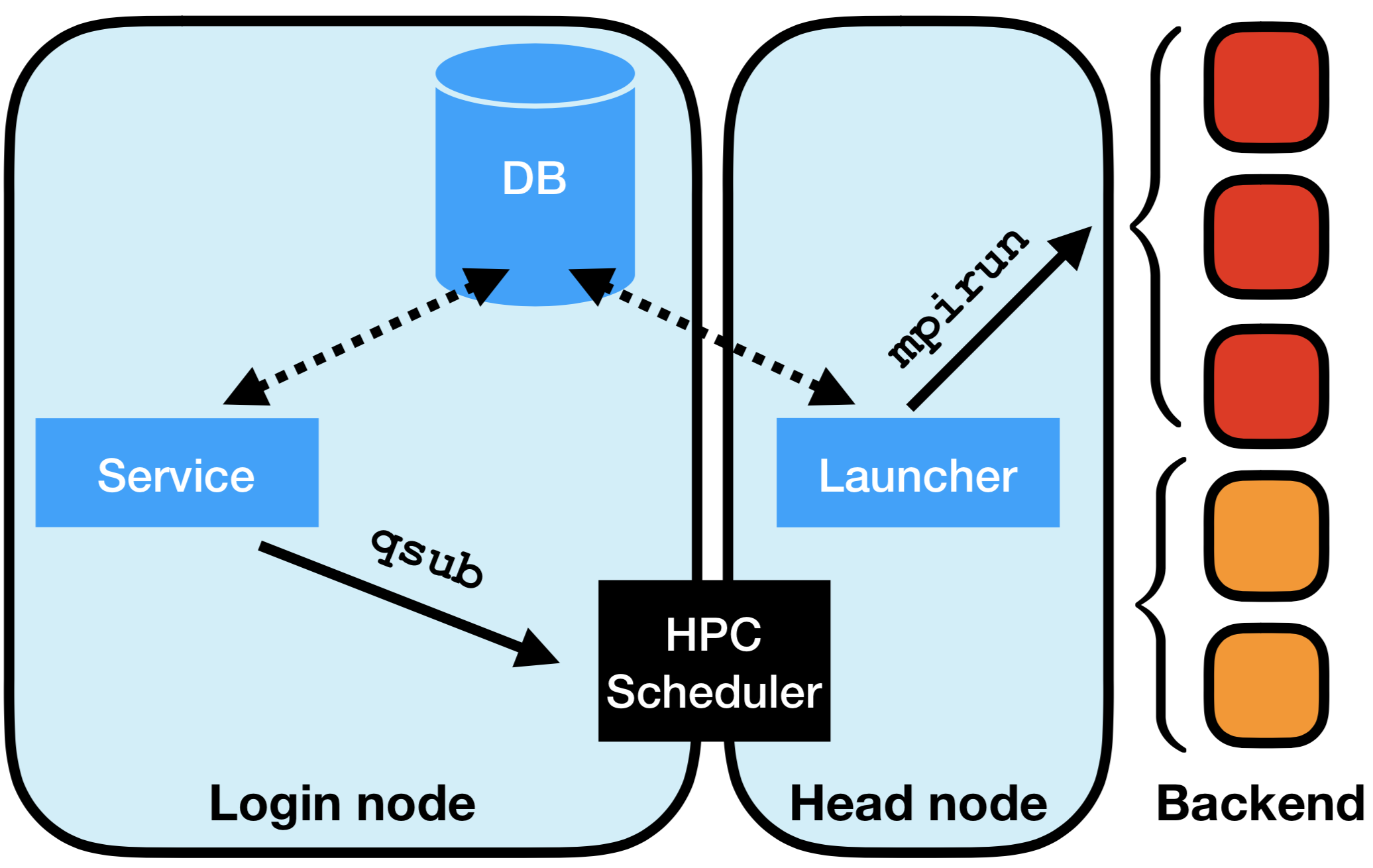}
    \caption{Schematic view of Balsam architecture in an HPC environment. Shown
    in blue are the Balsam database server, service, and launcher processes.
    The service runs on a login node, packs \balsamjobs{} for launcher-mediated
    execution, and submits them through a modular scheduler interface (e.g. Cobalt
    \texttt{qsub}) to a local queue.  When these jobs start on a designated
    head node, the launcher pulls \balsamjobs{} from the database and
    begins to dispatch tasks using a template for the local MPI implementation
    (e.g. \texttt{mpirun}).}
    \label{fig:balsam-overview}
\end{figure}

As an alternative to manually submitting launcher jobs, the user can choose to 
run the Balsam \textbf{service} as a background process
to manage scheduling in an automated fashion. In this case, the service packs
\balsamjobs{} into dynamically-sized requests for total compute nodes and wall
time, respecting a user-defined queueing policy and attempting to maximize
throughput (i.e. number of tasks finished per queued job). The job queued by
the Balsam service invokes the launcher exactly as if it were submitted
directly by a user.  In fact, launchers are almost entirely decoupled from the
service and from one another, so that it is possible to run several launcher
jobs concurrently to consume work from the same database.  When the service
submits a launcher job, however, it tags the \balsamjobs{} intended for execution
accordingly, and the launcher filters tasks by this tag in order to run its
pre-designated workload. Figure \ref{fig:balsam-overview} illustrates the
three-component architecture of Balsam running inside a typical supercomputer
infrastructure.

\subsection{Workflow and Data Model\label{sec:datamodel}}

A high-throughput, data-intensive workflow typically entails managing at least
thousands of tasks, executed across numerous batch jobs potentially spanning
months.  The \balsamjob{} database provides a stateful, persistent
representation of the job campaign to facilitate scheduling and execution.
Balsam leverages Django's object-relational mapper \cite{django} to provide a
simple Python object model for the \balsamjob{}.  The underlying PostgreSQL or
SQLite database is entirely abstracted from users, who interact with an
intuitive command-line interface or Python API to manage project databases or
add and modify tasks therein. 

\subsubsection{Command Line Interface}
In the absence of dependencies or data flow between tasks, Balsam provides
concise command line facilities to define tasks, invoke launcher execution, or
start the service to package tasks and throttle queue submission over time. 
This usage is illustrated in Listing
\ref{listing:addjob}.  Here, a Balsam database is initalized and populated with
100 independent tasks from the \texttt{bash} shell.  First, a new Balsam
database named \texttt{my-wf} is created in the current working directory with
\texttt{balsam init}.  Users may create separate databases for different
projects using this command, and context-switching is performed with the
\texttt{source balsamactivate} command, which sets the necessary local
environment variables and launches a database server process if necessary. The
\texttt{balsam app} command is invoked to register an executable with the
\texttt{run-sim} alias in the \texttt{ApplicationDefinition} table. Each entry
references an executable or Python script located on the filesystem, along with
optional pre- and post-processing scripts that can be used to create dynamic
workflows (discussed in section \ref{sec:dynamicwf}). Finally, several 4-node
MPI tasks are added to the database with \texttt{balsam job}. In this example,
each task runs an instance of the \texttt{run-sim} program, specifying a unique
input file that will be staged into the task-local working directory, which is
created by the Balsam launcher at runtime.

\begin{listing}[!t]
\centering
\small
    \begin{minted}{bash}
$ balsam init my-wf
$ source balsamactivate my-wf
$ balsam app --name=run-sim --exec=bin/sim.x
$ for i in {1..100};
  do
    balsam job --name=task$i --workflow=mini \
    --application=run-sim --num-nodes=4 \
    --ranks-per-node=16 --stage-in=inbox/$i.inp;
  done
    \end{minted}
    \caption{Inserting \balsamjobs{} from the shell\label{listing:addjob}}
\end{listing}

The \balsamjob{} table contains numerous fields to control how a task is
processed.  The minimum requirements, shown in Listing \ref{listing:addjob},
include a \texttt{name}, a \texttt{workflow} tag for grouping related tasks,
and an \texttt{application} referencing the task executable. The application
will run with MPI when either \texttt{num-nodes} or \texttt{ranks-per-node} is
set larger than 1, and there are fields to specify environment variables and
command line arguments. Wall time estimates for scheduling can be provided
with \texttt{wall-time-minutes}. A unique \balsamjob{} ID (UUID) is automatically
generated for each task and prevents naming collisions, while serving as the
primary key in the underlying database.

\subsubsection{DAG Workflows\label{sec:dagdefining}}
To define more complex workflows with data-dependencies between tasks, it is
possible to use the \texttt{balsam dep <parent-id> <child-id>} command to
create dependencies between parent and child tasks, which are then understood
in terms of the DAG structure where vertices represent tasks and directed edges
represent dependencies among them. It is often easier to express these
dependencies and build workflows programatically from a Python script using the
Balsam API, as illustrated in Listing \ref{listing:dagjobs}. This Listing
describes how one can programatically create a DAG. We omit several fields in
this example for brevity; nevertheless, the Listing is runnable, and Balsam
provides sensible default values for each missing \balsamjob{} field. 

\begin{listing}[!t]
\centering
\small
    \begin{minted}{python}
from balsam.launcher.dag import (
    add_job, add_dependency
)

A = add_job(name="A", application="generate")
B, C, D = [
           add_job(
             name=name, 
             application="simulate",
             input_files=name+".inp"
           )
           for name in "BCD"
          ]
E = add_job(name="E", application="reduce",
            input_files="*.out")

for job in B,C,D:
    add_dependency(A, job)
    add_dependency(job, E)
    \end{minted}
    \caption{Creating a DAG workflow from the Python API\label{listing:dagjobs}}
\end{listing}

\begin{figure}[!t]
    \centering
    \includegraphics[width=2.5in]{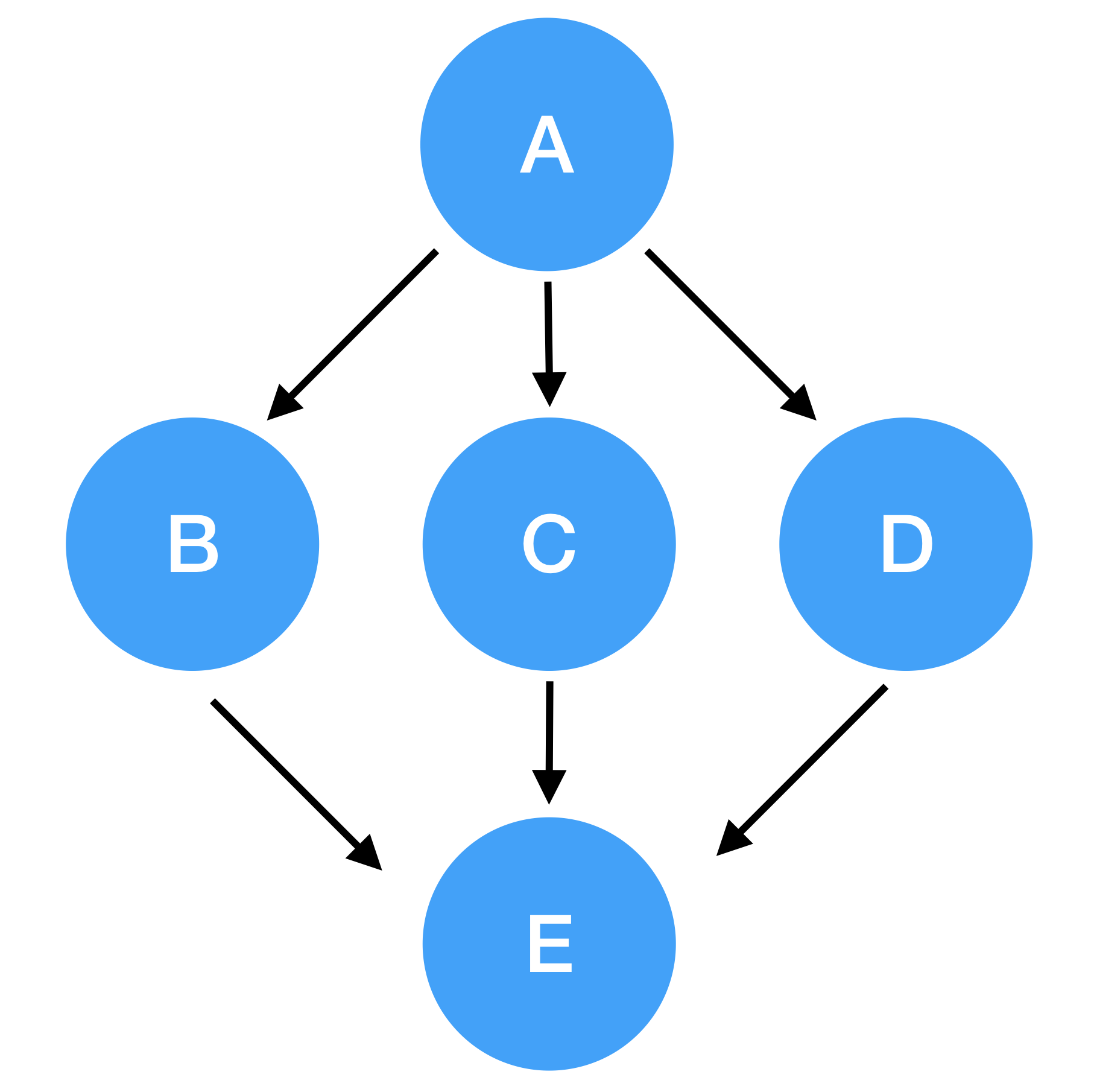}
    \caption{DAG corresponding to workflow defined in Listing \ref{listing:dagjobs} 
    \label{fig:dag}}
\end{figure}

Listing \ref{listing:dagjobs} creates five tasks in a "diamond" workflow
pattern (drawn in Figure \ref{fig:dag}). Parent job \texttt{A} runs the
\texttt{generate} application to produce several input files for the following
\texttt{simulate} stage. The parent fans out to three child jobs, \texttt{B},
\texttt{C}, and \texttt{D}, each of which runs a \texttt{simulate} instance on
one of the accordingly-named input files produced by \texttt{A}.  Finally, the \texttt{reduce}
application is run in task \texttt{E}, which depends on the \texttt{.out} files
of all three \texttt{simulate} applications.  The \texttt{input\_files} field
demonstrated here provides space-delimited filename patterns, such as \texttt{*.out}.
Files matching any of the patterns flow to a child \balsamjob{} from all of its
parents' working directories. This facilitates customizable dataflow from
parent to child tasks. When possible, symbolic links are created in the child
working directories to reduce unnecessary data movement in shared filesystems.
The \texttt{input\_files} option pertains only to data flow along DAG edges;
there are other options like \texttt{stage\_in} used in Listing
\ref{listing:addjob} that enable data movement from arbitrary locations with a
number of protocols (see section \ref{sec:transitions}).

\subsubsection{Task Processing and State Flow} 
The notion of \balsamjob{} state is at the heart of how workflows are processed
in Balsam.  The \texttt{state} field in the database tracks the status of each
task, from the moment of its creation until it is finished. Each \balsamjob{}
flows from one state to the next as it undergoes transitions by the launcher,
which groups tasks according to state to determine the necessary actions. The 
stateful, persistent representation of a workflow confers fault tolerance 
and allows execution across several queued batch jobs over time.

Tasks are easily listed according to
state with \texttt{balsam ls}, shown in Listing \ref{listing:ls} for the
previous sample DAG workflow. Balsam records a complete \texttt{state\_history}
for each task as well, with timestamps marking each transition and useful log
messages alongside them.  For instance, applications that return a nonzero
error code cause a \balsamjob{} to be advanced from \texttt{RUNNING} to
\texttt{RUN\_ERROR} and ultimately \texttt{FAILED} if no error-handling
mechanism is defined.  The error code and tail of the \texttt{stderr} stream
are recorded directly in \texttt{state\_history}, and a user can diagnose errors
at-a-glance by running \texttt{balsam ls --state=FAILED --history} to display the
logs of failed tasks.  

\begin{listing*}[!t]
\centering
\small

\begin{verbatim}
  $ balsam ls
                              job_id | name | workflow | application |            state
---------------------------------------------------------------------------------------
d487a785-3ff1-4702-aff9-d6f1f88dd795 | A    | sample   | generate    | JOB_FINISHED
94905135-b47d-439f-9561-6c16290112db | B    | sample   | simulate    | JOB_FINISHED
c04942d2-4926-4324-ad6e-b9c729d9f62b | C    | sample   | simulate    | RUNNING
19b130c3-50df-497c-a740-8c987c6b8e19 | D    | sample   | simulate    | RUNNING
15df7441-4fb9-4537-af96-5f91453b7f3a | E    | sample   | reduce      | AWAITING_PARENTS
  $
\end{verbatim}

\caption{Listing \balsamjobs{} from the shell\label{listing:ls}}
\end{listing*}

In addition to debugging workflows, the Balsam \texttt{state\_history} data captures 
sufficiently detailed timing information at the workflow level to gauge throughput and
supercomputer resource utilization over time. Convenience functions accessible from the
Python API, for instance, enable users to quickly generate a temporal profile of the
distribution of jobs across different states. This is particularly useful in carrying out
the sorts of efficiency analyses presented in section \ref{sec:deephyper}. 
A benefit of the Django ORM underlying Balsam is that power users can easily bypass
the Balsam API and work directly with the underlying \balsamjob{} model to
perform expressive, custom queries. For instance, ``\emph{find me all \texttt{SimX}
jobs that took at least 256 nodes and failed, then run them again}'' is as
simple as: \texttt{BalsamJob.objects.filter( \\
application="SimX", num\_nodes\_\_gte=256, \\
state="FAILED" \\
).update(state="RESTART\_READY")}.

\subsection{Launcher\label{sec:launcher}}
The Balsam launcher is the pilot application running on backend nodes to
process tasks in the database and carry out the state flow described in the
last section. Launcher instances are entirely decoupled from one another and
the Balsam service, and they can be invoked manually in a hand-written batch
job through the \texttt{balsam launcher} command, or scheduled automatically
via the service. Many launchers can run simultaneously to consume work from the
same database, potentially in batch jobs used from multiple users. The
underlying relational database is leveraged to ensure consistency in multi-user scenarios with
concurrent batch jobs, while providing an efficient means for storage and
retrieval of millions of tasks.

The launcher is invoked in one of two modes by setting the \texttt{--job-mode}
argument either to \texttt{serial} or \texttt{mpi}. In the former, only
single-node tasks that use applications without MPI are permitted to run; each
task application is forked on a backend compute node to run in isolation of the
network.  In the latter, tasks of any size are permitted, and each application
is launched independently under a separate \texttt{mpirun} (or equivalent)
system command. The \texttt{serial} mode is preferable in workflows entailing
hundreds of single-node tasks, because only a single MPI run process on the head
node is created, bypassing the potential resource burden of numerous \texttt{mpirun} processes. 
In all cases, the launcher runs continuously across the
alloted nodes and walltime, acquiring tasks from the database and mapping them
to idle resources in order to maximize concurrency and maintain load balance.
These methods are entirely flexibile with respect to the type of application,
permitting MPMD ensemble jobs without any modification to source code or
executables. Moreover, the \texttt{mpi} launcher naturally handles tasks
of mixed node counts and durations.

A key feature of all launcher modes is task-level fault tolerance, meaning that because tasks
run using isolated resources, software faults in one task do
not affect the launcher or other tasks running concurrently. 
The persistence of state in the
database also provides workflow-level resilience.  Launchers shutdown
gracefully and mark timed-out runs upon receiving termination signals or
running out of batch job walltime. Restarting a workflow is as simple as
running the launcher a second time. 

Finally, the launcher supports highly \emph{dynamic} workflows. In the course
of processing, \balsamjobs{} can create or kill other tasks according to
runtime logic. Users may even add or kill tasks manually from an interactive
shell on a separate login node. The launcher responds in near real-time,
stopping killed tasks mid-execution and proceeding to fill idle resources with
newly available tasks. In the following subsections, we provide some highlights 
of the launcher components and their implementation. The next section \ref{sec:dynamicwf} 
details how dynamic workflows can be created by attaching pre- or post-processing
scripts to certain applications.

\subsubsection{Transitions\label{sec:transitions}}
The launcher contains a \texttt{transitions} module responsible for carrying
out all of the pre- and post-execution \balsamjob{} state transitions.
Depending on the amount of I/O or pre/post-processing computation, the Balsam
configuration allows users to specify a number of transition processes that run
in parallel through the Python \texttt{multiprocessing} mechanism.  

The transitions module carries out a variety of pre- and post-execution functions.
First, the database is checked for newly created \balsamjobs{}; those without
dependencies are marked \texttt{READY} for processing, while those with pending
parents are marked \texttt{AWAITING\_PARENTS}. The \texttt{stage\_in}
transition creates the \balsamjob{}-unique working path, copies in remote or
local files, and ensures that for job dependencies, the requested files from
all parent tasks are visible to the present child task (as described in section
\ref{sec:dagdefining}.

The \texttt{preprocess} and \texttt{postprocess} transitions execute
user-defined preprocessing and postprocessing scripts, which are optionally
attached to a particular application in the \texttt{ApplicationDefinition}
entry.  These scripts, typically written in Python, run in the \balsamjob{}
working directory and encapsulate steps like generating input files
to an application and parsing the outputs for storage or further processing.  A
key feature in Balsam is that these scripts run in a special,
\balsamjob{}-aware environment that permits the calling code to get the current
job and its state.  We will show in \ref{sec:dynamicwf} that this permits users to
write dynamic workflows, in which a task programatically modifies the database
according to its outcome. 

\subsubsection{Serial Job Mode}
Some job campaigns entail a massive number of single-node jobs that make no use
of MPI or HPC network resources. In this case, the \texttt{balsam launcher
--job-mode=serial} command dispatches a \emph{fork-ensemble} MPI wrapper
process on the backend to handle all tasks that do not use MPI
(\texttt{num\_nodes} and \texttt{ranks\_per\_node} are equal to 1). The
advantage of this job mode is that only one MPI launch command is invoked on
the head node, and users can define ``packable'' \balsamjobs{} with
\texttt{node\_packing\_count} in order to run multiple tasks concurrently inside a
single node.

The serial job mode was highly optimized to minimize latency between task
injection and execution in Balsam. This is particularly important to the
performance of dynamic, asynchronous workflows, where new \balsamjobs{} are
created over time and run concurrently within a single resource allocation.  In
Section \ref{sec:deephyper}, we highlight the use of the serial job mode to
parallelize tens of thousands of single-node Tensorflow instances across 1024
nodes of the Theta leadership-class Cray XC40 system at the ALCF.  The same
workflow was run on the ALCF's Cooley visualization cluster with
\texttt{node\_packing\_count} set to 2, enabling each task to use one
accelerator of the NVIDIA Tesla K80 dual-GPU card.

\subsubsection{MPI Job Mode}
The \texttt{balsam launcher --job-mode=mpi} command invokes the Balsam launcher
on a head node, without any management processes running on the backend.  Here,
individual \balsamjobs{} are launched directly by issuing the appropriate
\texttt{mpirun} variant for the local MPI implementation. This mode of
execution is completely flexible with respect to the type of executables that
may run, but it may strain the local resource manager if several hundreds or
thousands of concurrent \texttt{mpirun} processes are used to run small (\textless 10 node)
parallel tasks (see discussion in Section \ref{sec:related}). 
Major advantages of this approach are total resource isolation
between \balsamjob{} tasks, conferring fault-tolerance as described in the
previous sections, and ease-of-use: running applications in any launch mode of
Balsam requires no source code modification or relinking of executables. The
latter point is often a serious bottleneck in WMS like Swift/T, discussed in
section \ref{sec:related}.

Much like the serial-mode launcher, tasks in this mode are dynamically mapped
to idle nodes for concurrent execution. Eligible \balsamjobs{} are sorted in
descending order of required nodes, so that the largest blocks of compute nodes
are allocated first.  This pattern follows the ``first-fit descending'' greedy
heuristic used for automatic scheduling by the Balsam service (Section
\ref{sec:service}).  As a result, when a user opts to automate job scheduling
with Balsam, the \balsamjob{} execution order and conurrency will approximately
match the intended schedule, provided that walltime estimates are
sufficiently accurate. On the other hand, the launcher naturally handles
inevitable departures from the ideal schedule, and load balance is maintained,
at least to first order, when tasks finish early or late, or unanticipated
errors arise in the workflow. 

\subsection{Dynamic Workflows \label{sec:dynamicwf}}
Another advantage of the Balsam launcher's real time task-pull behavior is that
\balsamjobs{} can be added or killed at any time. Even during a running
launcher batch job, a user can modify the database from a login shell to stop
certain tasks and restart them with more promising input parameters. 

The Balsam Python API, which was introduced in Section \ref{sec:dagdefining},
provides a workflow-aware environment in which tasks can modify the DAG
or otherwise manipulate the database during workflow execution.
If a \balsamjob{} application is written in Python itself, the task can interface
with the database while executing from the backend. A more general approach is
to write Python post-processing (or pre-processing) scripts that
run under the \texttt{transitions} module (Section \ref{sec:transitions}) and
manipulate the \balsamjob{} database based on the specific outcome of each task.

A task-aware environment is loaded upon importing the
\texttt{balsam.launcher.dag} module, providing workflow authors with the
\balsamjob{} object of the current execution context.  For instance, a
postprocessing script can be written to check for \texttt{RUN\_TIMEOUT} or
\texttt{RUN\_ERROR} states, read partial outputs, and take some action
(e.g. spawning new tasks) to handle these exceptional states.

An example of \balsamjob{} querying with the underlying Django ORM is given in
Listing \ref{listing:kill}, where a snippet of code checks for \texttt{RUNNING}
tasks with names containing \texttt{sim3}, and then uses \texttt{dag.kill} to
terminate each task and ensure that any child nodes in the DAG are marked
accordingly. This is a complete, runnable sample that could be invoked from within a
\balsamjob{} application or processing script.

\begin{listing}[!t]
\centering
\small
    \begin{minted}{python}
from balsam.launcher import dag

manager = dag.BalsamJob.objects
pending = manager.filter(name__contains="sim3", 
          state="RUNNING")
for sim in pending:
    dag.kill(sim, recursive=True)
    \end{minted}
    \caption{Dynamic \balsamjob{} kill\label{listing:kill}}
\end{listing}

\subsection{Balsam Service\label{sec:service}}
A scientific campaign can entail millions of calculations spanning months of
compute time on potentially multiple supercomputers with different job queueing
policies. In this case, different batch scheduler interfaces and limitations on
number of queued jobs require a cumbersome human effort to track workflow
execution and enqueue new jobs.  Unlike WMS with a DAG-centric view, the Balsam
service can be used as a simple, global abstraction of the local job scheduler.
Balsam throttles queue submission, while rounding small jobs into larger
ensembles in order to leverage the local queueing policies.

Central to the service is a user-defined \emph{queue policy}. Here, users
define the names of queues to which jobs may be submitted.  For each queue, a
limit on the number of concurrently queued jobs is given, along with a
dictionary mapping nonoverlapping node-count ranges to permitted walltime
ranges. A typical entry like \texttt{(128, 255) : (0.5, 3)} indicates that the
service may submit jobs requesting between 128 and 255 nodes, and the requested
walltime must lie between 30 minutes and 3 hours.  This allows Balsam to make
scheduling decisions within the bounds of local site policies; for instance,
larger jobs are typically permitted larger walltime limits in HPC facilities to
promote massively parallel applications. 

Invoking \texttt{balsam service} command launches a persistent background process
that runs on the user's behalf against the active Balsam database.
Jobs that are not already scheduled for launch or currently held by a launcher,
and which have yet to be processed, are eligible for scheduling. Balsam uses
greedy scheduling heuristics to pack many \balsamjobs{} efficiently into elastic
batch jobs. Deviations from the planned schedule are handled naturally by the 
Balsam launcher, and this process is robust to unexpected deletion of queued jobs,
requiring no user intervention.

The service employs a modular design, with a pluggable Scheduler interface
allowing Balsam to interoperate with a variety of local resource managers
(Cobalt, Torque, Condor, Slurm, etc\ldots). A template of the batch job script
is used to configure and invoke the launcher on the backend.  There is
virtually no interprocess communication between the service and launchers;
instead, shared state is captured in the database. Refer to the Balsam documentation
\cite{balsam-docs} for more detailed information on the job scheduling methods and 
design details.

\section{Applications\label{sec:applications}}
Balsam has been key to the success of several applications on ALCF systems. The
first of these was ATLAS event generation, for which Balsam was initially
developed. Here, we describe two applications that benefit specifically from
the present extensions to Balsam: the dynamic workflow model, launcher, and
scheduling service.  We first highlight DeepHyper, an application developed for
highly-parallel hyperparameter optimizations of deep neural networks. This
application leveraged the launcher to run model training and validation tasks
concurrently on up to 1024 nodes of the ALCF Theta supercomputer ($\sim 25\%$
of the machine) with fault tolerance and support for externally-signalled early
termination of tasks.  The second application showcases the efficacy of Balsam
in ensembles of MPI tasks, running a large set of \emph{ab initio} single-point
energy calculations with NWChem \cite{nwchem}.

\subsection{DeepHyper\label{sec:deephyper}}
We interfaced Balsam with the DeepHyper \cite{deephyper} package for
large-scale hyperparameter searches on deep neural networks. DeepHyper is a
Python framework that bridges user-defined benchmarks and hyperparameter
optimization methods through a generic interface. This architecture facilitates
testing arbitrary combinations of search methods and machine learning
benchmarks, while serving as an extensible repository of hyperparameter tuning
experiments.  Here, the Balsam launcher was used as the execution engine for a
variety of search experiments running on up to 1024 compute nodes of
Theta\cite{theta}, a leadership-class Cray XC40 system at the ALCF.

We first describe the DeepHyper-Balsam interface for training and evaluating ML
model configurations. We then highlight the performance of some asynchronous
search algorithms with the Balsam backend on both Theta and Cooley, a smaller
visualization and analysis cluster of 126 nodes with two GPUs per node.  We
show that the latency between job injection and execution is short enough that
Balsam poses no bottleneck to DeepHyper search at scale. DeepHyper-Balsam was
applied in a comprehensive hyperparameter optimization study using 3 million
Theta core-hours to-date \cite{deephyper}.

\subsubsection{Balsam Interface}
The DeepHyper \texttt{search} subpackage contains a number of optimization modules, including random search,
Bayesian optimization with surrogate models, evolutionary algorithms, and Hyperband \cite{hyperband}.
All search methods are implemented using the \texttt{Evaluator} three function interface, shown in Listing
\ref{evaluator}. All search modules use the \texttt{add\_eval\_batch} method to initiate new hyperparameter
evaluation tasks. This generic interface facilitates implementation of \emph{asynchronous} methods like random
search, in which which new hyperparameters can be sampled as soon as one or more evaluations have completed.
The corresponding method \texttt{get\_finished\_evals} is used to retrieve newly-finished hyperparameter
configurations along with the evaluated objective values.

\begin{listing}[!t]
\centering
\small
    \begin{minted}{python}
class Evaluator:
  def add_eval_batch(self, XX):
  '''Submit list of new cfgs XX'''
  def get_finished_evals(self):
  '''Query for newly completed evaluations
  and return results list: [(x, y)]'''
  def await_evals(self, XX):
  '''Block until evaluations are completed 
  for all cfgs in XX and return results list'''
    \end{minted}
    \caption{\texttt{Evaluator} interface}\label{evaluator}
\end{listing}

The \texttt{BalsamEvaluator} provides an implementation of this interface for
search execution on HPC resources, insulating search method developers from MPI
programming or other parallelization concerns.  This approach is also naturally
fault-tolerant; because Balsam marks failed tasks accordingly in the database,
DeepHyper is easily configured to either discard or assign a dummy objective
value (\texttt{sys.float\_info.max}) to failed hyperparameter evaluation tasks.
Most importantly, the \texttt{BalsamEvaluator} implementation incurred a
minimum of development overhead and made no references to any parallel
programming constructs.  Instead, the Django ORM was leveraged to simply insert
batches of evaluation tasks and query for successful and failed evaluations.
Listing \ref{amls-loop} illustrates how concisely search algorithms can be
expressed in this framework. After removing debugging and checkpointing
statements, the main loop of asynchronous Bayesian model-based optimization is
expressed in seven lines of Python. The \texttt{optimizer} object in this Listing
makes use of the ``ask-and-tell'' \texttt{scikit-optimize} interface
\cite{skopt}, and the \texttt{evaluator} is an instance of the
\texttt{BalsamEvaluator}.

\begin{listing}[!t]
\centering
\small
\begin{minted}{python}
for elapsed_str in timer:
    results = list(evaluator.get_finished_evals())
    if results:
        optimizer.tell(results)
        n = len(results)
        for batch in optimizer.ask(n):
            evaluator.add_eval_batch(batch)
\end{minted}
\caption{Asynchronous model-based search with \texttt{skopt}: main loop}\label{amls-loop}
\end{listing}

\subsubsection{Experimental Setup}
Balsam was installed into Anaconda environments configured for running
DeepHyper on both Theta and Cooley. Theta is an 11.69 petaflops Cray XC40 based
leadership-class supercomputing system at the ALCF, consisting of 4,392 nodes,
each with a 64-core Intel Xeon Phi processor.  Cooley is a 126 node cluster;
each node has 12 CPU cores and one NVIDIA Tesla K80 dual-GPU card. The
DeepHyper environment consisted of Intel Python 3.6.3, Tensorflow 1.3.1
\cite{abadi2016tensorflow}, Keras 2.0.9 \cite{chollet2015keras}, and
scikit-optimize 0.4 \cite{skopt}.  The Balsam launcher is host-aware and
configured to produce the appropriate \texttt{aprun} or \texttt{mpirun} startup
commands on Theta MOM nodes and Cooley head nodes, respectively. Experiments
were initated directly from hand-written Cobalt submission scripts to run the
Balsam launcher. On Theta, the
Balsam launcher was run in serial task mode, with one worker rank per node. Two
ranks-per-node were used on Cooley, enabling concurrent evaluation of a
separate model on each GPU.  To quantify the impact of database backend, we
compared the performance of random search with two Django backends: PostgreSQL
\cite{postgres} and SQLite \cite{sqlite}. With PostgreSQL backend, the launcher
is able to leverage transactions and reduce load on the server by batching
database updates. The SQLite backend, which required a custom server
implementation to serialize writes, did not utilize Django support for
transactions.

\subsubsection{Scaling Tests}
We present results using the random search (RS) method of DeepHyper on two
benchmarks: \texttt{rnn2} is a memory network obtained using a recurrent neural
network trained on the bAbI dataset for question-and-answer systems, and
\texttt{cifar10cnn} is a convolutional neural network benchmark for classifying
images in the cifar10 dataset. The baseline codes for these benchmarks were
obtained from the examples directory of the Keras github repository
\cite{chollet2015keras}. The RS method was chosen for the speed at which new
hyperparameters can be sampled from a uniform distribution. By minimizing time
spent in the the search algorithm itself, we can clearly gauge overheads of the
DeepHyper+Balsam framework and evaluate whether the system poses bottlenecks at
scale.

\begin{figure}[!t]
    \centering
    \includegraphics[width=3.0in]{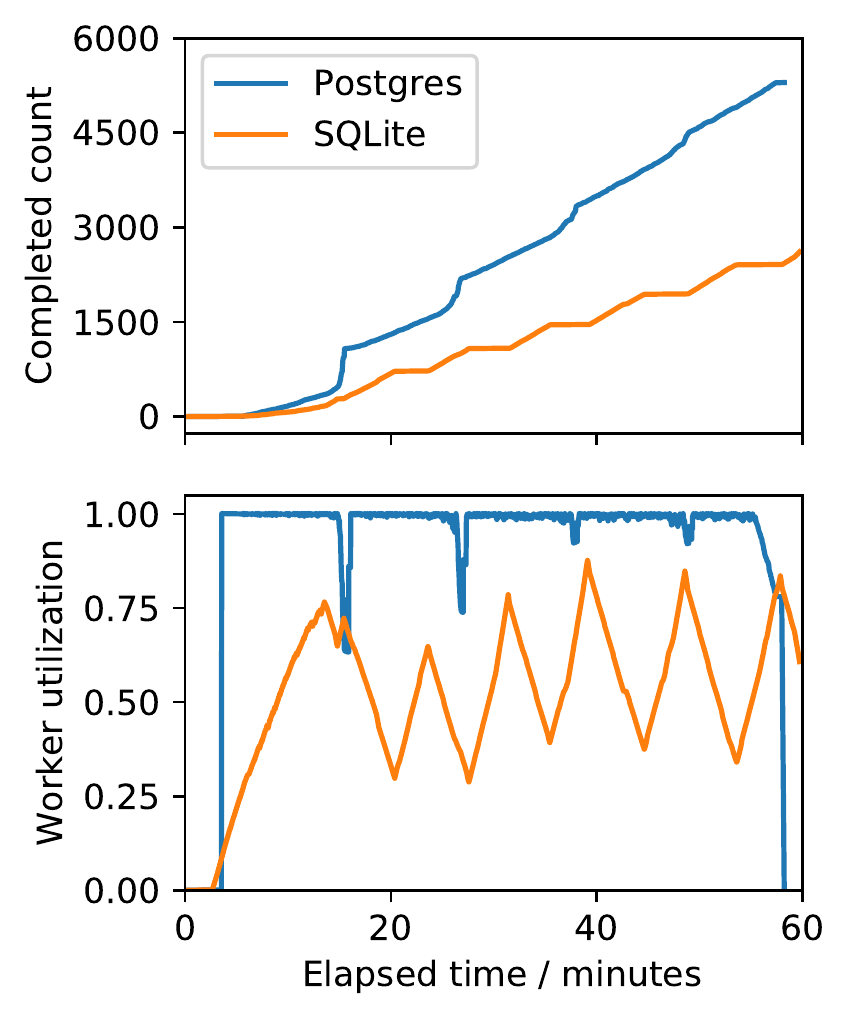}
    \caption{DeepHyper random search (RS) for the \texttt{rnn2} benchmark
    running on 1024 Theta nodes. The top panel charts the number of completed
    hyperparameter evaluation tasks against elapsed search time. The bottom
    panel plots worker utilization, which is computed as the fraction of worker
    nodes actively running a task at any given time. Curves drawn in blue and orange
    highlight Balsam performance with PostgreSQL and SQLite backends, respectively. }
    \label{fig:deephyper-pg}
\end{figure}

Nine hyperparameters were scanned for \texttt{rnn2}: number of training epochs,
number of hidden layers, number of units per hidden layer, activation function,
batch size, dropout, optimizer type, learning rate, and RNN type; details on
the hyperparameter space definition are available in a recent publication
\cite{deephyper}.  In Figure \ref{fig:deephyper-pg}, we evaluate the
performance of RS for this benchmark running on 1024 Theta nodes. The impact of
the Django backend database is borne out in the difference between the traces
corresponding to PostgreSQL and SQLite.

The top panel shows that the throughput is nearly twice as high when the
PostgreSQL backend is used with Balsam, as compared with the SQLIte backend.  Given the median \texttt{rnn2} task
runtime of 621 seconds on a single Theta node, one expects 5.80 tasks completed
per node-hour in the absence of overheads due to Balsam and DeepHyper. For the
PostgreSQL RS experiment, we found 5328 completed tasks in 54.31 minutes (measured
from the time of initial task creation to completion of the final task), which
amounts to nearly ideal efficiency at 5.75 tasks per node-hour. In fairness,
this timing omits 4 minutes in the job prior to initialization of the RS and 2
minutes spent on incomplete tasks prior to walltime expiration. If the 5328
tasks are divided by the full 1024 node-hour allocation, this still amounts to
5.20 tasks completed per node-hour, or 90\% efficiency. We observe that about 3
minutes of startup time elapsed in loading the Balsam launcher; this time spent
in loading Python is not intrinsic to Balsam \emph{per se}.

In 128 node experiments, Balsam overheads are so small that is little need for
concern with database access patterns, and there is hardly an observable difference
between PostgreSQL and SQLite backends. At 1024 nodes, however, these overheads
become large enough that the SQLite backend is unable to sustain high worker
utilization. The SQLite utilization curve fluctuates between 30\% and 80\%,
which manifests in the lower throughput seen in the top panel of Figure
\ref{fig:deephyper-pg}. A discussion of the performance issues at scale is
provided in the Appendix (Section \ref{sec:appendix}). 

Figure \ref{fig:rs-scaling} demonstrates the scalability of DeepHyper+Balsam RS
with the PostgreSQL backend up to 1024 Theta nodes. Here, as the number of
nodes is increased, DeepHyper RS provides accordingly many \texttt{rnn2}
configurations for concurrent, asynchronous evaluation. As expected from the
preceding discussion, since worker utilization remains high even at 1024 nodes,
the task throughput scales efficiently. We observed a 7.64-fold increase in
throughput on scaling from 128 to 1024 nodes, amounting to 96\% weak-scaling
efficiency.

\begin{figure}[!t]
    \centering
    \includegraphics[width=3.0in]{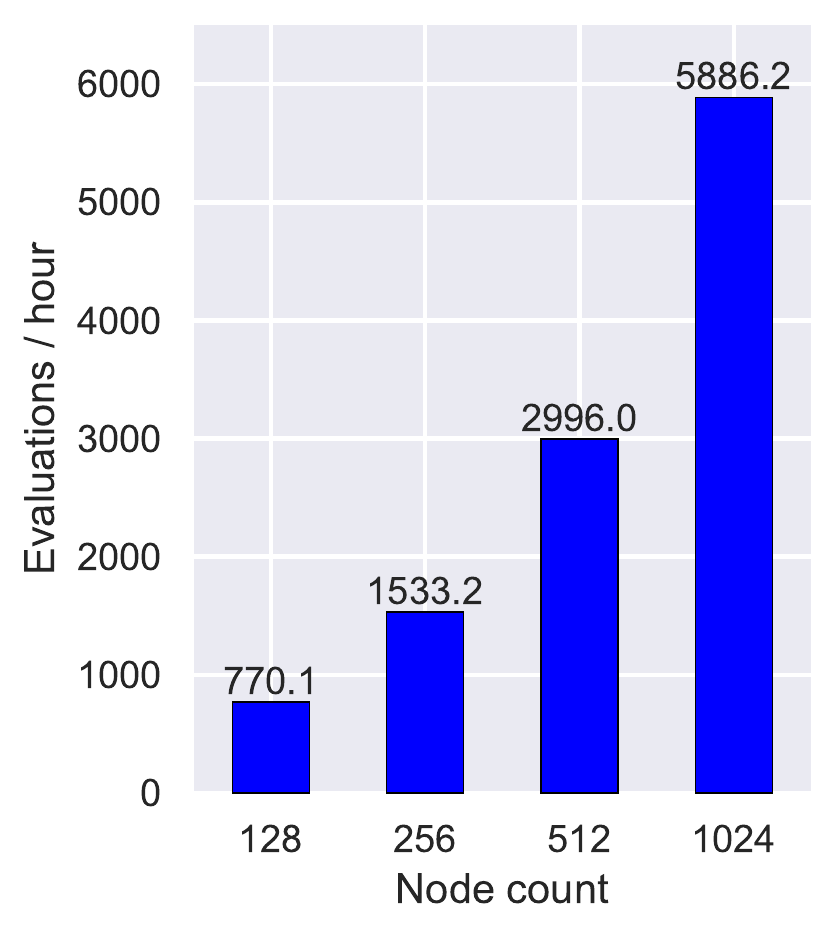}
    \caption{DeepHyper random search (RS) for the \texttt{rnn2} benchmark
    running on 128 to 1024 Theta nodes, using the PostgreSQL Balsam backend. The throughput, defined as number of
    completed \texttt{rnn2} evaluations-per-hour, is charted against compute
    node count.  While each job was nominally 60 minutes, the number of
    completed evaluations is divided by the elapsed time between creation of
    the first task and completion of the final task, which is about 54.5 minutes in
    each case.}
    \label{fig:rs-scaling}
\end{figure}

\begin{figure}[!t]
    \centering
    \includegraphics[width=3.0in]{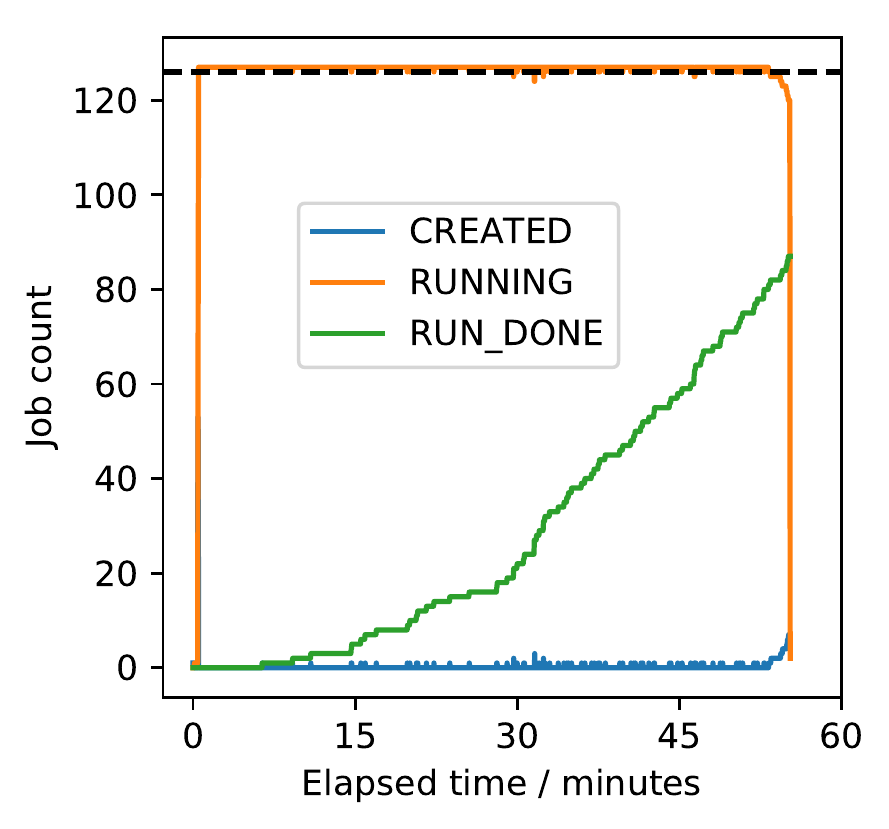}
    \caption{DeepHyper asynchronous model-based search (using random forest) for the \texttt{cifar10cnn} benchmark
    running on 64 Cooley nodes, with 2 Balsam workers per node. This panel shows a characterstic workflow
    execution profile which can be obtained from the Balsam API \texttt{service.models.process\_job\_times}.
    The \texttt{CREATED} trace shows the number of newly injected tasks from DeepHyper that have not
    been processed. The \texttt{RUNNING} curve gives the number of actively running tasks at any given time,
    from which worker utilization can be inferred. The \texttt{RUN\_DONE} curve shows the number of completed
    tasks.}
    \label{fig:deephyper-cooley}
\end{figure}

Finally, we show another temporal profile in Figure \ref{fig:deephyper-cooley} tracking
the performance of a DeepHyper+Balsam run on 64 nodes of Cooley (2 concurrent tasks
per-node; one per-GPU). Results are shown for the \texttt{cifar10cnn} benchmark
with an asynchronous Bayesian model-based search algorithm and SQLite backend.
In the Balsam context, this search method has an identical workflow pattern to
random search.  The traces presented in
this figure were directly obtained from the Balsam Python API; the call to
\texttt{service.models.process\_job\_times()} infers the number of tasks in each
state versus time from stored provenance data.
The orange trace corresponding to \texttt{RUNNING} tasks can simply be normalized
by the number of workers to obtain the worker utilization (as in Figure
\ref{fig:deephyper-pg}).  With a significantly longer runtimes per-task and
smaller number of workers than the previous example, we find the SQLite backend
is sufficient to sustain 100\% utilization.

\subsection{Quantum Chemistry: Potential Energy Scan}

Next, we present a common workflow in materials science and computational
chemistry that highlights launching of many concurrent MPI jobs.  The problem
is to scan a molecular structure across a range of internal
coordinates and compute the electronic energy for each structure under a given
model chemistry. The computed values are then assembled to construct an
electronic potential energy surface (PES) for the molecule.

\subsubsection{Configuration}
For this example, we take the symmetric water molecule H$_2$O and sample the O--H
bond length and H--O--H bond angle at 40 values for each coordinate, resulting in a total of
1600 geometries.  We compute the electronic energy at the SCS-MP2/aug-cc-pVTZ level
of theory using NWChem \cite{nwchem} version 6.6 compiled for the Theta KNL nodes.
Each energy calculation was parallelized across 64 MPI ranks on 2 Theta nodes.
A short Python script was used to scan the geometry and populate the Balsam database
with an NWChem task for each set of coordinates. The workflow was scheduled in a single
128-node Theta job, running the launcher in \texttt{mpi} job mode.

\subsubsection{Results}
The NWChem tasks ranged from 8 to 30 seconds long, with a mean task execution
time of 11 seconds.  All 1600 tasks were completed in a span of 9 minutes and
56 seconds, measured from launcher startup to completion of the last task.
This amounts to a throughput of about 2.7 tasks per second, \emph{including} the
overheads of a stage-in transition for copying input files to each task's
working directory, as well as the delay of the MPI launch command. In the
context of typical HPC workflows, these are very short-running tasks that
expose Balsam overheads.  Still, Balsam does not pose a signifcant bottleneck
here, and the overhead is expected to be negligible for more compute-intensive
workloads. In fact, Balsam significantly outperforms typical
recommendations for hand-written ensemble batch jobs on Theta, where only one
\texttt{aprun} is launched per second.

\begin{figure}[!t]
    \centering
    \includegraphics[width=3.0in]{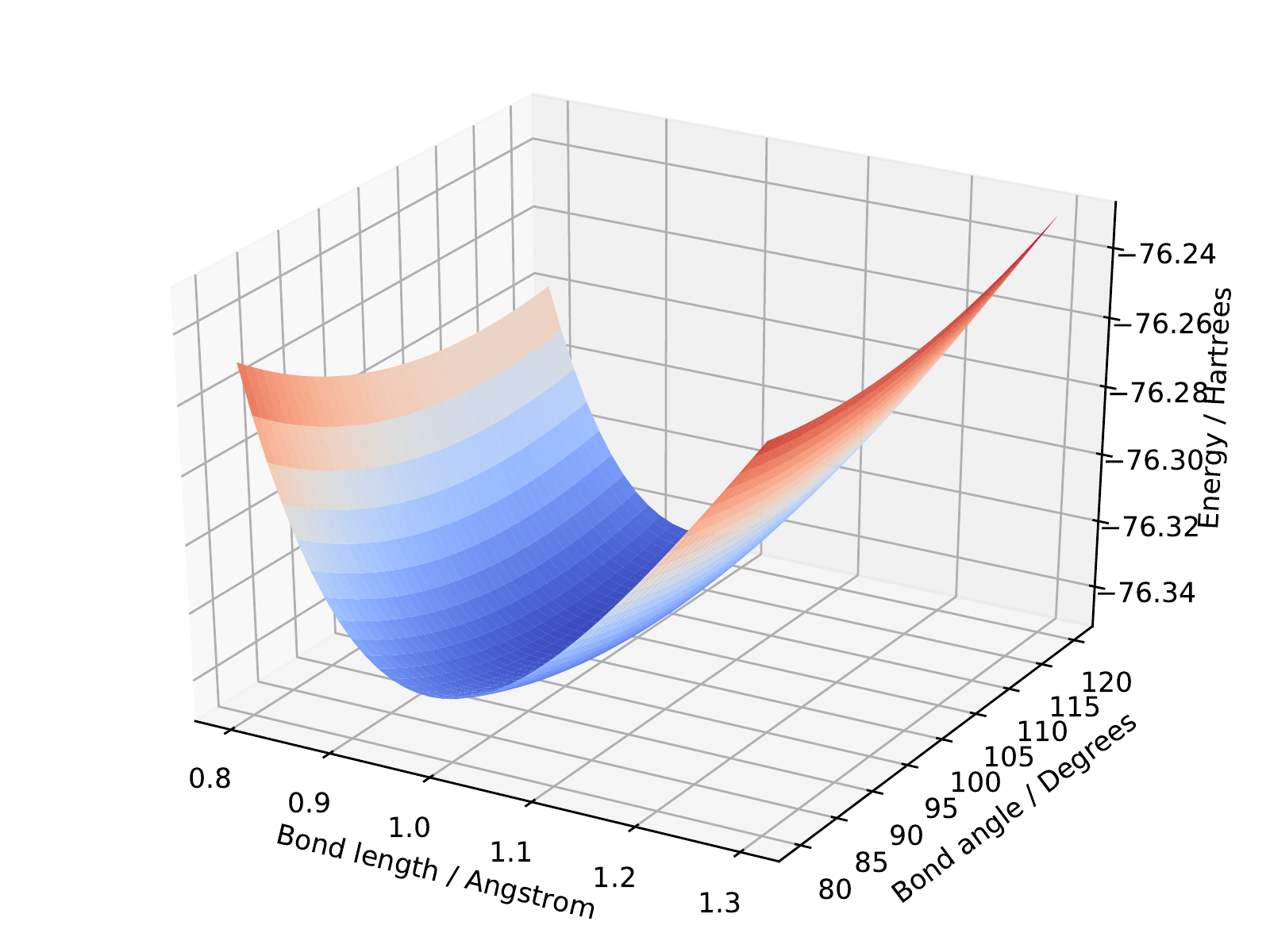}
    \caption{SCS-MP2/aug-cc-pVTZ potential energy surface of the water molecule in \emph{C2v} symmetry.
    The electronic energy was computed with a Balsam/NWChem workflow at each of 1600 geometries, and is 
    plotted in atomic units against O--H bond length and H--O--H bond angle.}
    \label{fig:water-pes}
\end{figure}

The \balsamjob{} description was used to tag each task according to the
coordinate values. A trivial Python script with the \texttt{dag} API was then
used to collect the calculation outputs to assemble the water PES from all
1600 points.  The outcome is shown in Figure \ref{fig:water-pes}, which plots a
smooth 2D surface over the PES values computed at each bond length and angle
value.  We emphasize the ease-of-use of Balsam in this scenario; the scripts to
generate the entire workflow and produce this Figure required under 60 lines of
Python and minimal effort on the user's behalf. More importantly, an
existing NWChem binary was plugged into the workflow with a single
\texttt{ApplicationDefinition} and no modifications whatsoever. Balsam effectively
load-balanced the 1600 2-node tasks across 128 total nodes, achieving excellent 
throughput and storing provenance data which made PES generation trivial.

\section{Summary and Future Work}

Balsam's loosely coupled
service--database--launcher architecture provides a flexible, yet very
accessible framework for managing large computational campaigns with a minimum
of development or cognitive overhead. The service provides elastic scheduling
of ensemble jobs, enabling users to leverage supercomputer scheduling policies
to increase workflow throughput. The launcher enables efficient execution of
heterogenous workloads both concurrently and sequentially across compute
allocations, with task-level fault tolerance and support for highly dynamic
workflows.  Detailed documentation and examples for using Balsam can be found
online \cite{balsam-docs}.

The benefits of automated job scheduling and ensemble-packing are useful only
insofar as the runtime of tasks can be estimated with some precision. Towards
this goal, we will investigate the use of automated performance models to
utilize measured task runtimes for predicting the duration of pending tasks in
the database.  High-throughput campaigns with repetitive usage
patterns can provide a wealth of actionable performance data for the Balsam
service, especially if users provide the mapping from task input to
features determining computational complexity. A general framework for updating
application performance models and predicting runtimes with uncertainty quantification
could be used to assign resources to tasks and optimize the scheduling process.

Several users have requested the ability to ``co-locate'' a chain of sequential
tasks in the workflow, such that the tasks are grouped for execution on the
same set of processors to exploit local disks or shared memory. Strategies for
scheduling and execution of co-located tasks in the Balsam launcher will be
explored in upcoming work. We will also continue to investigate alternative
launcher execution modes, particularly for concurrent execution of many small
MPI tasks, where conventional methods impose limits on scalability. A primary
challenge here lies in portability across MPI implementations.

The version of Balsam described here will be reintegrated with the HPC edge
service \cite{hpc-edge17} for remote workflow management.  The web interface for 
monitoring workflows will reflect the support for organizing workflows as task graphs 
and should ultimately provide users with a centralized source of provenance data for an
entire computational campaign.

\section*{Acknowledgment}
This work was funded in part and used resources of the Argonne Leadership Computing Facility at Argonne
National Laboratory, a DOE Office of Science User Facility supported under Contract DE-AC02-06CH11357.

The submitted manuscript has been created by UChicago Argonne, LLC, Operator of Argonne National Laboratory (``Argonne"). Argonne, a U.S. Department of Energy Office of Science laboratory, is operated under Contract No. DE-AC02-06CH11357. The U.S. Government retains for itself, and others acting on its behalf, a paid-up nonexclusive, irrevocable worldwide license in said article to reproduce, prepare derivative works, distribute copies to the public, and perform publicly and display publicly, by or on behalf of the Government.


\bibliographystyle{IEEEtran}
\bibliography{bib/sw.bib}
\section{Appendix: Artifact Description \label{sec:appendix}}
\subsection{Setup and Portability}
As of this writing, Balsam requires Python 3.6, \texttt{mpi4py} version 2 or
newer, and PostgreSQL to be installed on the system. These prerequisites are
easily satisified on most modern platforms with a package management system
(e.g. Anaconda) and freely-distributed Postgres binaries. The \texttt{balsam
init} and \texttt{source balsamactivate} commands encapsulate all details of
database creation and server management, requiring no configuration on the part
of the user for most use cases. A default Balsam database and user
configuration file are created in \texttt{.balsam} under the user's home
directory.  Here, the service queueing policies and various constraints on
launcher execution can be adjusted from the default values if necessary.

The modular design of Balsam allows it to run on diverse HPC platforms and computing
clusters by providing the system-appropriate plug-ins. For instance, the service
uses a \texttt{Scheduler} interface to query the local batch job queue and submit
new launcher jobs. A number of scheduler plug-ins are provided (Cobalt,
Condor, Slurm, Torque), and can be extended with minimal effort.  The launcher
reads a scheduler-specific job environment to determine the available compute nodes
and remaining walltime; these are easily defined in the scheduler plug-in as well.
Finally, MPI launch commands are rendered using a template appropriate to the
MPI runtime environment (Cray aprun, OpenMPI, and MPICH, etc\ldots). The user needs only 
specify which template is appropriate to the system, or define a template if it is
not already supported.

The \texttt{serial} launcher job mode is only supported on platforms that
support the \texttt{fork} system call. Moreover, applications built with MPI
but invoked to run as serial programs can be problematic on some
supercomputers, where the call to \texttt{MPI\_Init()} from the backend causes
an error in PMI. There is ongoing work to develop effective and portable launcher
job modes that circumvent these limitations.

\subsection{DeepHyper Experiments}
The efficient task throughput of DeepHyper RS with Balsam and the PostgreSQL
backend can be understood in terms of the high average worker utilization,
shown in the bottom panel of Figure \ref{fig:deephyper-pg}. The utilization,
which can be inferred directly from the Balsam database post-run, is the
fraction of workers (number of Theta nodes minus 2) executing a
\texttt{BalsamJob} at a given point in time. With the PostgreSQL backend,
utilization is nearly pegged at 100\%, and Balsam poses no significant
bottleneck to the search workflow. Periodic drops in utilization last about one
minute and occur when a large number of \texttt{rnn2} tasks with similar
runtime finish together in a short interval. The following cycle occurs until
utilization is restored to 100\%:
\begin{enumerate}
    \item {all workers ranks that finished a task send a \texttt{RUN\_DONE} message to the Balsam launcher
        master process;}
    \item {the launcher batches all $N$ of the received \texttt{RUN\_DONE} messages in a 1-second window, and performs
        the corresponding batch database update;}
    \item {the DeepHyper RS code, which queries the database for finished tasks
        every 2-seconds (via the \texttt{BalsamEvaluator}), receives a list of $N$
        newly-finished tasks;}
    \item {the DeepHyper RS draws $N$ new hyperparameter configurations at random and
        inserts this batch of $N$ new tasks into the Balsam database, again via the
        \texttt{Evaluator} API;}
    \item {the launcher refreshes its job cache, receiving these $N$ new tasks. A task is sent to each
        idle worker until all idle workers have resumed execution of \texttt{rnn2} tasks, and utilization
        is restored to maximum.}
\end{enumerate}
When task completion is highly staggered in time, the number of finished tasks
$N$ in a given interval remains low and the fluctuations below 100\% utilization
are very small and short-lived. In the DeepHyper \texttt{rnn2} benchmark,
however, the runtime of different hyperparameter configurations forms a
strongly-peaked distribution, causing $N$ to be large for several consecutive iterations.
This effectively incurs some communication overhead between DeepHyper and
Balsam \emph{through} the shared database.

The primary reason for the observed slowdown with SQLite at scale (Figure
\ref{fig:deephyper-pg}) is that the standard Python adapter to the client
library did not support concurrency, and we were forced to implement a
serialization mechanism to intercept database updates occuring from different
processes. In doing so, we sacrificed Django's transactional support, and
database updates incurred a cost proportional to the number of updated rows,
which is clearly non-scalable.  On the other hand, the Balsam launcher running
against the PostgreSQL backend optimizes database access patterns by pooling
changes in state and grouping updates into large but short-lived transactions.
In this way, the number of database transactions remains small and constant
with respect to increasing number of worker nodes or task throughput.

Twelve tasks in the search at 1024 nodes experienced a segmentation fault in
the \texttt{fork} system call on Theta compute nodes; the workflow was
unaffected by these faults.  We found that setting the environment variable
\texttt{MPICH\_GNI\_FORK\_MODE=FULLCOPY} solved the issue of these sporadic
faults, but caused the median \texttt{rnn2} task time to increase from 621 to
781 seconds.  Specifically, the distribution of elapsed times for network
\emph{training} acquired a heavy tail, with 90th percentile training time
increasing from 614 to 1787 seconds. Python module loading and \texttt{rnn2}
preprocessing times were unaffected. There is ongoing dialogue with Cray to
identify the cause of performance degradation and search for a better strategy
to invoke \texttt{fork} in Theta ensemble jobs. In any case, Balsam provided
strong task-level fault tolerance: the \texttt{BalsamJob} experiencing a fault
is marked accordingly, and DeepHyper provides the next evaluation task to the
worker node with minimal interruption in the workflow.

\end{document}